\documentclass[aps,prl,twocolumn,superscriptaddress]{revtex4-1}
\usepackage{hyperref}
\usepackage{graphicx,color}
\usepackage[small,centerlast]{caption}
\usepackage{import}
\usepackage{siunitx}

\usepackage[resetlabels]{multibib}
\newcites{SI}{nada}

\definecolor{mygrey}{rgb}{0.5,0.5,0.5}
\definecolor{mygreen}{rgb}{0,0.5,0}
\definecolor{mylightgreen}{rgb}{0,0.75,0}
\definecolor{myblue}{rgb}{0,0,0.75}
\definecolor{mymagenta}{cmyk}{.2,1,0,0.12}
\definecolor{myred}{rgb}{0.95,0,0}
\definecolor{myorange}{rgb}{1,.5,0}

\newcommand{\Sph}{S_{\rm ph}}
\newcommand{\Sat}{S_{\rm at}}
\newcommand{\Rbef}{$^{85}$Rb}

\newcommand{\PRLsection}[1]{\noindent{\textit{#1} ---}}
\newcommand{\Intens}{{\cal I}}
\newcommand{\INucl}{I}

\newcommand{\Aeff}{A_{\rm eff}}

\newcommand{\Lcell}{L_{\rm cell}}
\newcommand{\QE}{Q}

\usepackage{draftcopy}
\usepackage{amsmath,
,chngpage,mathtools,subfigure}

\begin{document}

\title{Squeezed-light spin noise spectroscopy}

\newcommand{\ICFO}
{
\affiliation{ICFO-Institut de Ciencies Fotoniques, The Barcelona Institute of Science and Technology, 08860 Castelldefels (Barcelona), Spain}}
\newcommand{\ICREA}
{
\affiliation{ICREA -- Instituci\'o Catalana de Recerca i Estudis
Avan\c{c}ats, 08015 Barcelona, Spain}
}
\newcommand{\ECNU}
{
\affiliation{East China Normal University, Shanghai 200062,
China}
}

\author{Vito Giovanni Lucivero}
\email{vito-giovanni.lucivero@icfo.es}
\ICFO
\author{Ricardo Jim\'{e}nez-Mart\'{i}nez}
\ICFO
\author{Jia Kong}
\ICFO
\author{Morgan W. Mitchell}
\ICFO
\ICREA


\date{May 1, 2016}

\newcommand{\SINotMath}[2]{#1 #2}
\renewcommand{\micro}{[micro]}
\renewcommand{\SI}[2]{\ifmmode \mbox{\SINotMath{#1}{#2}} \else \SINotMath{#1}{#2} \fi}

\begin{abstract}
We report quantum enhancement of Faraday rotation spin noise spectroscopy by
polarization squeezing of the probe beam. Using natural abundance Rb in \SI{100}{Torr}of N$_2$
buffer gas, and squeezed light from a sub-threshold optical parametric oscillator stabilized \SI{20}{GHz}to the blue of the D$_1$ resonance, we observe that an input squeezing of \SI{3.0}{dB} improves the signal-to-noise ratio by \SI{1.5}{dB} to \SI{2.6}{dB} over the combined (power)$\otimes$(number density) ranges
($\SI{0.5}{mW}$ to $\SI{4.0}{mW}$)$\otimes$($\SI{1.5}{$\times 10^{12}$ cm$^{-3}$}$ to
$\SI{1.3}{$\times 10^{13}$ cm$^{-3}$}$), covering the ranges used in optimized spin noise spectroscopy experiments. We also show that squeezing improves the trade-off between statistical sensitivity and broadening effects, a previously unobserved quantum advantage.
\end{abstract}

\maketitle

The presence of intrinsic fluctuations of a spin system in
thermal equilibrium was first predicted by Bloch \cite{Bloch1946}
and experimentally demonstrated in the $1980$'s by Aleksandrov and
Zapasskii \cite{Zap1981}. In the last decade, ``spin noise
spectroscopy'' (SNS) has emerged as a powerful technique for
determining physical properties of an {\em unperturbed} spin
system from its noise power spectrum
\cite{Zapasskii2013,HAbner2014}. SNS has allowed measurement of
g-factors, nuclear spin, isotope abundance ratios and relaxation
rates of alkali atoms
\cite{Crooker2004,Katsoprinakis2007}, g-factors, relaxation times
and doping concentration of electrons in semiconductors
\cite{Oestreich2005,RAmer2009,Huang2011,Berski2013,Poltavtsev2014}
and localized holes in quantum dot ensembles
\cite{Crooker2010,Li2012} including single hole spin
detection \cite{Dahbashi2014}. Recently, SNS has been used to study complex optical
transitions and broadening processes \cite{Zapasskii2013a,Yang2014}, coherent phenomena beyond linear
response \cite{Glasenapp2014} and cross-correlations of
heterogeneous spin systems \cite{Dellis2014,Roy2015}.

Spin noise has been measured with nuclear magnetic resonance
\cite{Sleator1985,Reim1986} and magnetic force microscopy
\cite{Rugar2004,Budakian2005,Degen2007}, but the most sensitive
and widely used detection technique is Faraday rotation (FR)
\cite{Zap1981,Crooker2004,Katsoprinakis2007}, in which the spin
noise is mapped onto the polarization of an
off-resonant probe. In FR-SNS, spin noise near the Larmor
frequency competes with quantum noise \cite{Sorensen1998} of the
detected photons, i.e., the optical shot noise. The main figure
of merit is $\eta$, the peak power spectral density (PSD) due
to spin noise over the PSD due to shot noise, called ``signal strength'' \cite{MAller2010} or
the ``signal-to-noise ratio'' (SNR). Reported SNR for single-pass
atomic ensembles ranges from $0$ dB to $13$ dB
\cite{Crooker2004,Shah2010}, and up to $21$ dB in atomic
multi-pass cells \cite{Li2011}.   Due to weaker coupling to the
probe beam, reported SNR ranges from $-50$ dB  to $-20$ dB  in
semiconductor systems (See Table 1 in \cite{MAller2010}). Several works
have studied how to improve the polarimetric sensitivity
\cite{Glasenapp2013} or to cancel technical noise sources
\cite{Oestreich2005,RAmer2009,Berski2013}, but
without altering the fundamental tradeoff
between sensitivity and broadening processes \cite{Glasenapp2013}.
%
%

For small optical power $P$ and atomic density $n$, SNR is linear in each:
$\eta \propto nP$.  At higher values, light scattering and atomic collisions
broaden the spin noise resonances, and thus introduce systematic errors in measurements, e.g. of relaxation rates, that are derived from the SNS linewidth \cite{Crooker2004,Katsoprinakis2007,Oestreich2005}. This trade-off between statistical sensitivity and line broadening is a fundamental limitation of the technique, with its origin in quantum noise properties of the atomic and optical parts of the system.

Here we work in a high-density regime, with atomic number densities up to $n \sim \SI{\num{e13}}{cm$^{-3}$}$, covering the range of recent experiments with optimized atomic instruments \cite{Crooker2004,Katsoprinakis2007,Shah2010,Dellis2014}.  Earlier studies in this regime have observed non-trivial interactions between optical quantum noise and nonlinear magneto-optical rotation (NMOR) of a on-resonance probe \cite{Novikova2015} including increased measurement noise as a result of input squeezing for densities above $n \approx  \SI{\num{2e11}}{cm$^{-3}$}$  \cite{Horrom2012}.  It is thus not obvious that squeezing will improve a high-density Faraday rotation measurement \cite{Zapasskii2013,Budker2007}, as it does for lower densities \cite{Sorensen1998,Wolfgramm2010}. In contrast, here we observe that squeezing does in fact improve both the signal to noise ratio and the sensitivity/line broadening trade-off in SNS, over the full practical range of the technique. It is worth noting that we work with an un-polarized atomic ensemble and off-resonant probing, as required for the non-perturbative SNS technique. This may explain the difference between our results and prior experiments \cite{Horrom2012,Novikova2015}.

\newcommand{\VPD}{V_{\rm DPD}}

\PRLsection{Theory}
As described in detail in the Appendix, FR optical
probing gives a signal proportional to the on-axis projection of
the collective spin of a group of atoms in thermal equilibrium.  The collective spin precesses in
response to external magnetic fields and experiences a stochastic motion as required
by the fluctuation-dissipation theorem \cite{Zap1981,MAller2010}.
%
For rubidium, which has two isotopes $^{85}$Rb and $^{87}$Rb, and shot-noise
limited detection \cite{Lucivero2014}, the power spectrum of the FR
signal is given by a double Lorentzian function:
\begin{equation}
      S(\nu) = \Sph + \sum_{i\in\{85,87\}} \Sat^{(i)}\frac{(\Delta\nu_{i}/2)^2}{(\nu-\nu_L^{(i)})^2+(\Delta\nu_{i}/2)^2},
    \label{Eq:fitfunction}
\end{equation}
where $S_{\rm ph} \propto P \xi^2$ is the (frequency-independent)
shot-noise contribution at power $P$ and $\xi^2$ is the squeezing factor. The
sum on $i$ is over atomic mass, $\nu_L^{(i)}$ and $\Delta\nu_{i}$
are the (linear) Larmor frequency and FWHM width, respectively.
$\Sat^{(i)}$ is the height of Lorenzian spin noise contribution. The SNR is:
\begin{equation}
\eta_i \equiv \frac{\Sat^{(i)}}{\Sph} =  \frac{P}{\hbar \omega} \frac{4 \QE}{\xi^2} \frac{\sigma_0^2}{\Aeff^2}
\frac{\kappa_i^2 {n_i \Lcell \Aeff} }{\Gamma},
\label{SNR}
\end{equation}
where $\hbar \omega$ is the photon energy, $\QE$ is the quantum efficiency, $\sigma_0$ is the resonant optical cross-section of the collision-broadened optical transition, $\Aeff$ is the effective beam area, $L_{\rm cell}$ is the vapor cell length, $n_i$ is the number density and $\kappa_i^2$ is a coupling that depends on light detuning and angular momentum quantum number. $\Gamma=1/T_{2}$ is the spin relaxation rate, related to the FWHM of the noise spectrum  by $\Delta \nu = 1/(\pi T_{2})$, and given by \cite{Jimenez2012}:
\begin{equation}
    \frac{1}{T_{2}} = \Gamma_{0} + \alpha n + \beta P
    \label{Eq:T2main}
\end{equation}
where $\Gamma_0$ is the unperturbed line-width, $n = n_{85}+n_{87}$ is the total atomic density, and $\alpha$ and $\beta$ are collisional broadening and power-broadening factors, respectively.

From Eq. (\ref{SNR}) we see that it is in principle possible to increase $\eta$ by increasing
either the optical probe power or the atomic density. However
both of these actions result in additional broadening of the
linewidth $\Delta\nu$. As one main use of SNS is to measure relaxation
processes in an unperturbed spin system \cite{Crooker2004,Katsoprinakis2007}, this additional broadening represents a systematic shift of the measured variable (the linewidth) \cite{Glasenapp2014}. On the other hand, Eqs. (\ref{SNR})  and (\ref{Eq:T2main}) predict that squeezing boosts $\eta$ without additional shifts, providing a quantum advantage irrespective of the other experimental parameters.

\begin{figure}[t]
\centering
\hspace{-0.4cm}
{\includegraphics[width=\columnwidth]{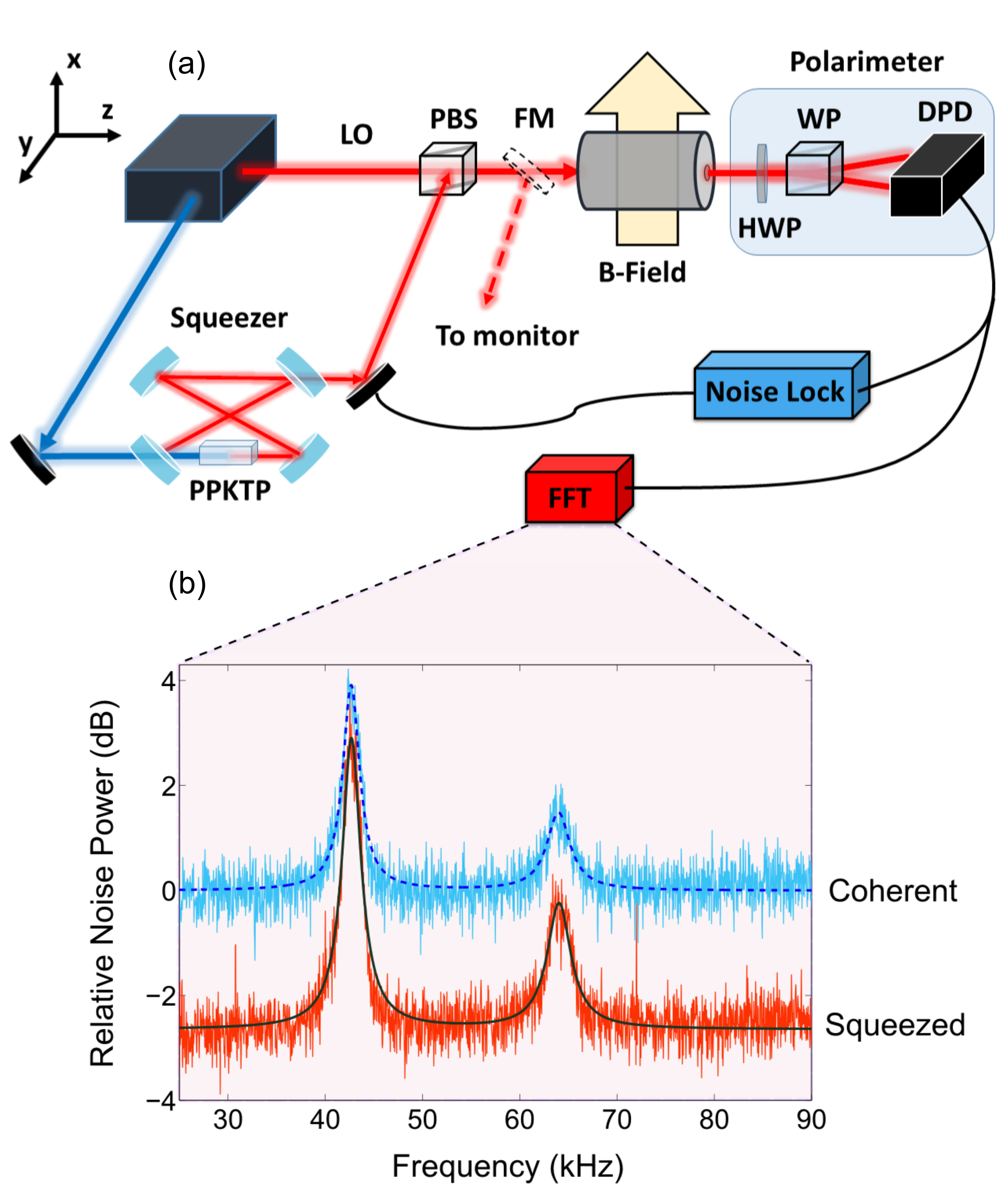}}
\caption{\textbf{Squeezed-light spin noise spectroscopy.}  \textbf{(a) Experimental schematic}. LO - local
oscillator, PBS - polarizing beam splitter, DPD - differential
photo detector, FM - flip mirror, HWP - half wave-plate, WP - Wollaston prism, FFT -
fast Fourier transform analyzer.
\textbf{(b) SNS Spectra}. Averaged spin noise spectra at $T=90^{\circ}$ ($n=\SI{2.4}{$\times 10^{12}$ cm$^{-3}$}$) acquired with coherent probe, cyan (light gray), and polarization squeezed probe, red (dark gray), respectively.  The spectra shown are averages of 10 spectra, each representing \SI{0.5}{s}of acquisition organized into bins of width $10$ Hz. \SI{0}{dB} on the power scale corresponds to \SI{-95.57}{dBV/Hz} at the detector. Dashed and continuous smooth curves show fits by Eq. (\ref{Eq:fitfunction}) to the coherent and squeezed spectra, respectively. Optical power $P=\SI{2.5}{mW}$, magnetic field $B_x = \SI{5.6}{$\mu$T}$.} \label{setup}
\end{figure}

\PRLsection{Experiment} The experimental setup is shown schematically in
Fig.~\ref{setup} (a). A polarization-squeezed probe beam is generated as in  \cite{Grangier1987}
by combining local oscillator (LO) laser light with orthogonally polarized squeezed vacuum using a
parametric oscillator described in \cite{PredojevicPRA2008}, cavity
locking system as in \cite{Beduini2014}, and a quantum noise lock (to ensure the measured Stokes component is the squeezed component) described in
\cite{McKenzie2005,Wolfgramm2010}. The
probe frequency is stabilized to \SI{20}{GHz} to the blue of the \Rbef~D$_1$ unshifted line using
the system of \cite{Kong2015}. Adjusting the LO power
changes the probe power $P$ without changing the degree of squeezing.
To perform conventional FR-SNS with the coherent LO
\cite{Zapasskii2013} we simply turn off the squeezer.
The effective size of the probe beam, as defined in the Appendix, is \SI{0.054}{cm$^{2}$}.

The atomic system consists of {an ovenized} cylindrical vapor cell of
length $L_{\rm cell} = \SI{3}{cm}$ and diameter $d=\SI{1}{cm}$, with natural
isotopic abundance Rb and \SI{100}{Torr} of N$_2$ buffer gas.  Density is controlled
by oven temperature and calibrated by absorption at 20 GHz detuning. We alternate five-second acquisition periods with electrical heating of the oven. Before the cell, measured squeezing at a sideband frequency of \SI{40}{kHz} is \SI{3.0}{dB}, while squeezing after the cell ranges from  \SI{2.6}{dB} to  \SI{1.5}{dB} at the highest density. These numbers are consistent with expected loss of squeezing due to absorption of the off-resonance probe (See Appendix).

We apply a transverse DC field $B_x = \SI{5.6}{$\mu$T}$ and
minimize the gradient $\partial B_x/\partial z$ by minimizing the width
of the SNS resonances. The oven and coils are inside four layers of high-permeability
magnetic shielding.  We detect the probe beam with a polarimeter consisting of
a half-waveplate, Wollaston prism and differential
photodetector (DPD). The output is recorded by a 24-bit digitizer with
\SI{200}{kHz} sampling rate and a PC computes the power spectrum.

In Fig.~\ref{setup} (b) we show typical spectra using coherent and polarization
squeezed probes. As expected from Eq. (\ref{Eq:fitfunction}) we
observe the two atomic noise contributions from $^{85}$Rb and
$^{87}$Rb centered at Larmor frequencies $\nu_L^{85}$ and
$\nu_L^{87}$ above a uniform shot noise background. Squeezing
reduces the shot noise level without evident change to the spin noise
contribution, resulting into a SNR improvement.

\begin{figure}[h!]
 \centering
\hspace{-0.4cm}
{\includegraphics[width=0.9\columnwidth]{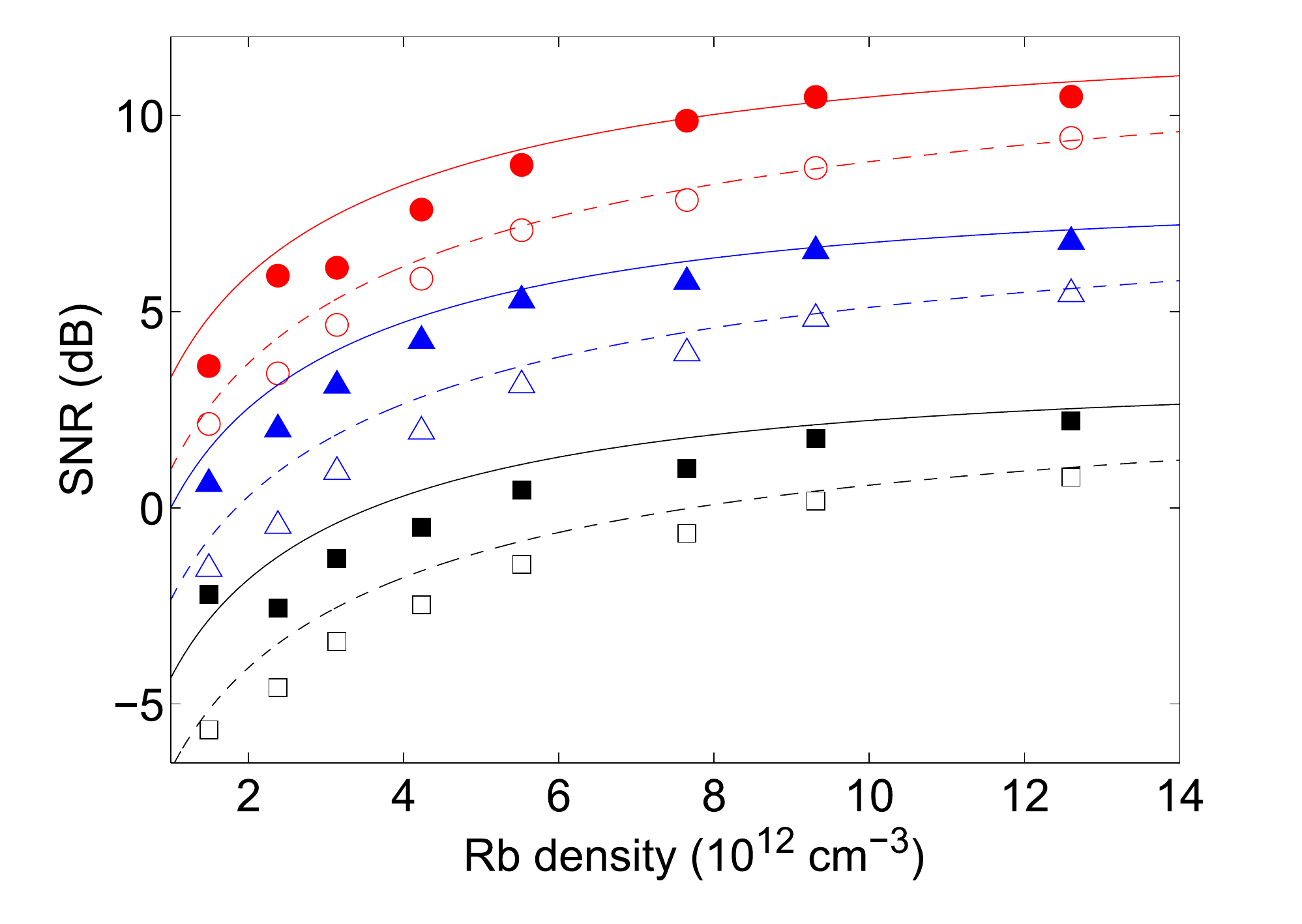}}
\caption{\textbf{SNR Enhancement}. SNR $\eta$ versus atomic density for
coherent probes (empty markers) and polarization
squeezed probes (filled markers), respectively. We show three optical powers of $P=\SI{0.5}{mW}$ (squares), $P=\SI{1.5}{mW}$ (triangles) and $P=\SI{4}{mW}$ (circles). Predicted $\eta$ from Eq. (\ref{SNR}) is plotted for coherent (dashed lines) and squeezed (continuous lines) probing, taking into account the reduction of squeezing versus density due to absorption at the probe frequency. The discrepancy between theoretical curves and experimental data is due to uncertainty on density estimation. (See Appendix)}
\label{SNRvsDens}
\end{figure}

\textit{Data analysis and results.}
At any given optical power and atomic density we acquire $100$ individual spin noise spectra as described in Fig. \ref{setup} (b)  and fit them with Eq. (\ref{Eq:fitfunction}) to obtain the parameters $S_{\rm ph}$ and $S_{\rm at}^{(i)}$, $\nu_L^{(i)}$, $\Delta\nu_{(i)}$, and the derived
$\eta = S_{\rm at}^{(i)}/S_{\rm ph}$, for $i \in \{85,87\}$. Due to imperfect stability of the quantum noise lock, it is necessary to reject about \SI{10}{percent}of the traces, identified by the condition $\chi \equiv\int d\nu \, S(\nu) > 1.03 \bar{\chi}$, where the integral is taken over a featureless window between $80-90$ kHz, not included in the fitting, and $\bar{\chi}$ is the average of $\chi$ over the $\rm 100$ spectra.

In Fig.~\ref{SNRvsDens} we show measured $^{85}$Rb SNR versus $n$,
the Rb number density, for coherent and squeezed probes at
$P=\SI{500}{$\mu$W}$ and $P=\SI{4}{mW}$, which bound our
investigated power range, and at the intermediate
$P=\SI{1.5}{mW}$. The temperature range is $T=\SI{85}{$^\circ$C}$ -- $\SI{120}{$^\circ$C}$.
As expected from Eq. (\ref{SNR}), the SNR increases with both increasing density and increasing power.
Squeezing enhances the SNR by a factor ranging from $2.6$ dB at $n
= \SI{1.5}{$\times 10^{12}$ cm$^{-3}$}$ to $1.5$ dB at $n =
\SI{1.3}{$\times 10^{13}$ cm$^{-3}$}$, with the difference due to
greater absorption at higher density. This latter number is higher than the densities used in other alkali SNS works \cite{Crooker2004,Katsoprinakis2007,Dellis2014}, and at this density we observe collisional broadening $\alpha n/(2 \pi) = \SI{760}{Hz}$,  larger than $\Gamma_0/(2\pi) = \SI{501}{Hz}$. We are thus in a regime where a feature of interest (the linewidth) is already strongly disturbed. Analogous observations apply to the investigated probe power range. In this sense the regime we investigate is fully practical for SNS applications \cite{Zapasskii2013,HAbner2014}.  Moreover, the model described in the Appendix shows that the benefit of squeezing extends until the optical absorption becomes strong and the squeezing is lost. The theoretical SNR for squeezed and coherent probes (curves of Fig.~\ref{SNRvsDens}) converge at a density of $n \approx 1.3\times10^{14}$ cm$^{-3}$ (temperature larger than $T=160^{\circ}$), much above the investigated and practical range of interest. This is our first main result: polarization squeezing significantly improves the SNR of SNS over the full practical range of power and density without any detrimental effect.

\begin{figure}[t]
 \centering
\vspace{0.4cm}
{\includegraphics[width=0.9\columnwidth]{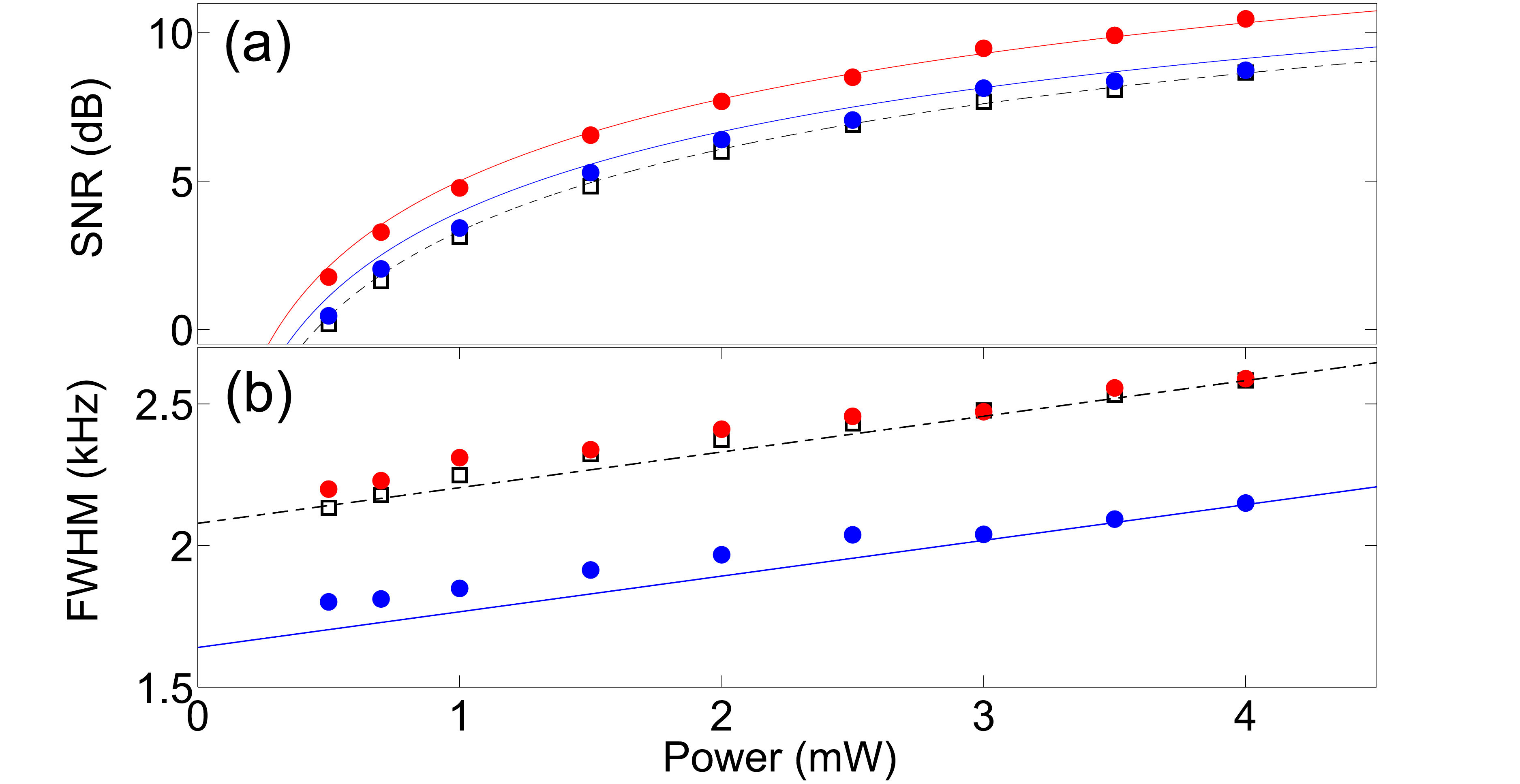}}
\caption{\textbf{Collisional broadening reduction} \textbf{(a)} SNR
$\eta$ versus optical power for coherent (empty black squares) and polarization
squeezed (filled red circles) probing at $n = \SI{0.9}{$\times 10^{13}$ cm$^{-3}$}$ and just for squeezed probe (filled blue circles) at lower density $n = \SI{0.5}{$\times 10^{13}$ cm$^{-3}$}$. Theoretical SNR from Eq. (\ref{SNR}) is shown for coherent (dashed lines) and squeezed (continuous lines) probing. \textbf{(b)} FWHM linewidth
$\Delta\nu_{85}$ vs probe power for the same conditions of (a). Theoretical FHWM widths from Eq. (\ref{Eq:T2main}) are plotted for coherent (dashed line) and squeezed (continuous line) probing, respectively.}
\label{SNRvsPower}
\end{figure}

Furthermore,  in
Fig.~\ref{SNRvsPower} we show $^{85}$Rb SNR $\eta_{85}$ and FWHM linewidth $\Delta\nu_{85}$ versus optical power for three different experimental situations:
$n = \SI{0.5}{$\times 10^{13}$ cm$^{-3}$}$ (squeezed only) and, with roughly twice the density, at $n = \SI{0.9}{$\times 10^{13}$ cm$^{-3}$}$  (coherent and squeezed). At the higher density, squeezing improves the SNR with respect to the coherent probe
without significantly changing the linewidth.    At the lower density, squeezing gives the same SNR as the coherent probe gives with the higher density, but with significantly less perturbation of the linewidth. These behaviors are observed over the full investigated power range.

\begin{figure}[h!]
 \centering
\hspace{-0.4cm}
{\includegraphics[width=0.9\columnwidth]{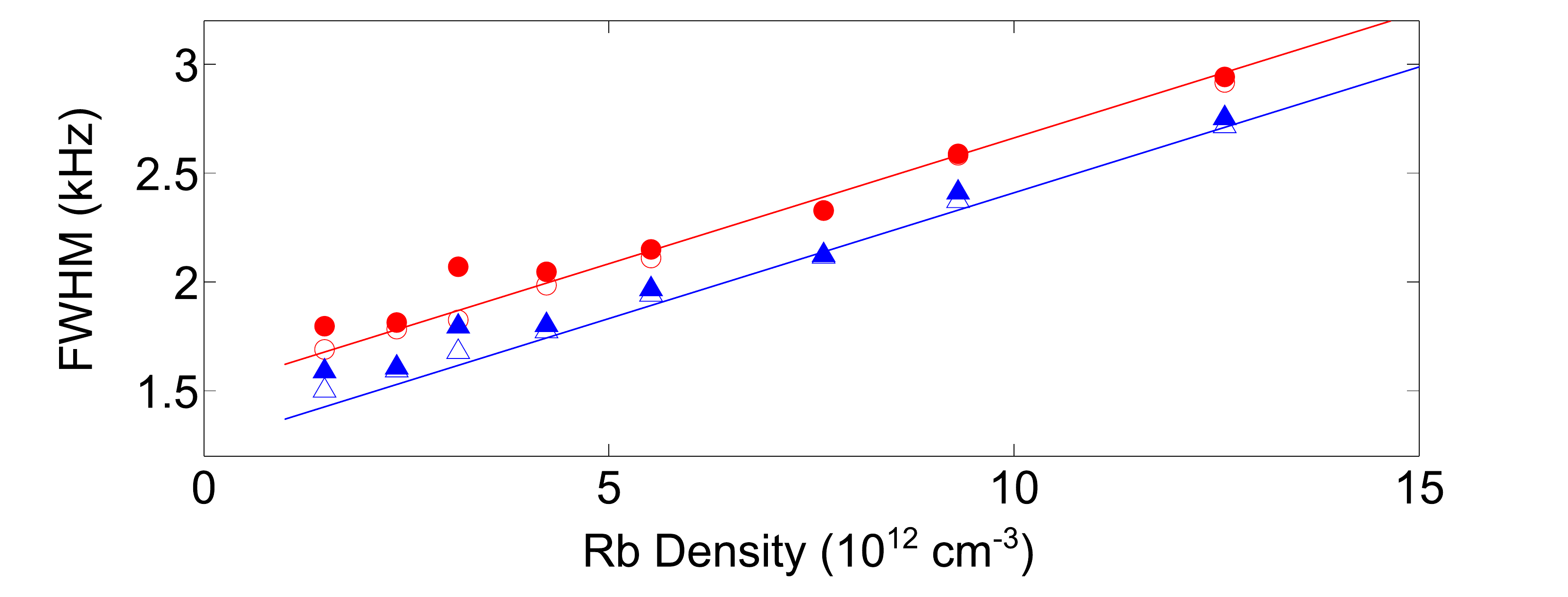}}
\caption{\textbf{Power broadening reduction} FWHM linewidth
$\Delta\nu_{85}$ vs Rb density for coherent (empty markers) and polarization
squeezed (filled markers) probing at optical powers of $P=\SI{4}{mW}$ (red circles) and $P=\SI{2}{mW}$ (blue triangles). Theoretical FHWM widths from Eq. (\ref{Eq:T2main}) are plotted.}
\label{FWHMvsDens}
\end{figure}

A similar behavior occurs if we compare SNR and linewidth versus atomic density at different probe power levels. Fig.~\ref{SNRvsPower} (a) already shows that squeezing allows us to get the same classical SNR by using about half of the power, resulting in a reduced power broadening. For completeness in Fig.~\ref{FWHMvsDens} we show the FWHM linewidth versus Rb density for $P=2$ mW and $P=4$ mW where the SNR with squeezing at $P=2$ mW is equal to the SNR with coherent probing at $P=4$ mW. The linewidth reduction with power is smaller than that seen Fig.~\ref{SNRvsPower} (b), because the collisional broadening is greater than the broadening due to probe scattering and, in these experimental conditions, dominates the broadening.  In Figs.~\ref{SNRvsPower} and \ref{FWHMvsDens} we show our second main result: squeezing can reduce broadening effects at no cost to figures of merit such as statistical sensitivity.

\PRLsection{Conclusions and outlook}
We have studied the application of polarization-squeezed light to spin noise spectroscopy of
atomic ensembles over the full practical range of density and probe power.  We observe that
squeezing improves the signal-to-noise ratio by an amount comparable to the applied squeezing, in
contrast to prior experiments \cite{Horrom2012} that showed the opposite behavior:
increased measurement noise due to squeezing above a critical density.  We demonstrate that by using a crystal-based squeezer and off-resonant probing of an un-polarized ensemble, differently from \cite{Horrom2012,Novikova2015}, optical and atomic quantum noise add incoherently without any coupling. Moreover, squeezing
improves the trade-off between statistical noise and line broadening by giving
performance not available with classical probes at any power level.

Our results provide clear evidence that squeezing can improve Faraday-rotation-based SNS measurements, with a broad range of applications in atomic and solid state physics \cite{Zapasskii2013,HAbner2014}. This advantage over the full practical parameter range for SNS is promising for similar advantages in high-performance magnetometers, gravimeters, and clocks.

\section*{Acknowledgements}

We thank Federica A. Beduini for the knowledge transfer on laser system and polarization squeezing source.
Work supported by the Spanish MINECO projects MAGO (Ref. FIS2011-23520) and EPEC (FIS2014-62181-EXP), Catalan 2014-SGR-1295, by the European Research Council project AQUMET, Horizon 2020 FET Proactive project QUIC and by Fundaci\'{o} Privada CELLEX.

\section*{APPENDIX}
In what follows we present expressions for the shot-noise background, $\Sph$, and the height of the Lorentzian spin noise,  $\Sat^{(i)}$, appearing in the function used to fit the power spectral density (PSD) of the polarimeter output as described in the main text. Using these expressions we obtain the SNR $\eta_{i} = \Sat^{(i)}/\Sph$ (i.e. Eq.(2) of the main text) which is used to estimate $\eta$ for $^{85}$Rb as a function of probe light power and density (see solid and dashed lined in Figs. 2 and 3). In generating these estimates we have used the parameters quoted in Table \ref{table:parameters}.

\newcommand{\Eph}{E_{\rm ph}}
\newcommand{\omegalight}{\omega_{\rm light}}
\renewcommand{\omegalight}{\omega}
\newcommand{\nulight}{\nu_{\rm light}}
\newcommand{\var}{{\rm var}~}

\subsection{Detector signal}

In our experiments we analyze $S(\nu)$,  the power spectral density (PSD) of $V_{\rm DPD}$, the output voltage of the polarimeter shown in Fig.~1, expressed in \si{\volt \squared \per {\hertz} }.  Because the scalar signal is acquired by combining polarization rotation information over $\cal{A}$, the area of the beam, we write the signal as

\begin{equation}
    V_{\rm DPD}(t) =  2 G \Re [ P \Theta_{\rm FR}(t) + P_{\rm SN}(t) ],
    \label{Eq:VPD}
\end{equation}
where $G =10^6$ V/A is the transimpedance gain, $P =\int_{\cal A} dxdy\, \Intens(x,y)$ is the total power of the beam reaching the detector with intensity $\Intens(x,y)$, and $\Theta_{\rm FR} \ll 1$ is the Faraday rotation (FR) angle as defined in Eq.~(\ref{Eq:ThetaFFromFz}) below. $\Re =  Q q/\Eph$ is the detector responsivity, where $Q$ denotes the quantum-efficiency of the detector,  $\Eph= \hbar \omegalight = 2.49 10^{-19}$J is the photon energy at \SI{795}{nm}, and $q = 1.6 10^{-19}$C. $P_{\rm SN}$ is a white-noise component due to shot noise, which we now compute.

\subsection{Photon shot-noise} \label{section: photon-shot noise}

The contribution from photon shot-noise to $S(\nu)$ is given by

\begin{equation}
    S_{\rm ph}  = 2G^2 q (\Re P) \xi^{2},
    \label{Eq:Sph}
\end{equation}
\noindent where $\xi^2$ represents the light-squeezing parameter.

\begin{figure}[!htb]
    \centering
    \includegraphics[scale=0.65]{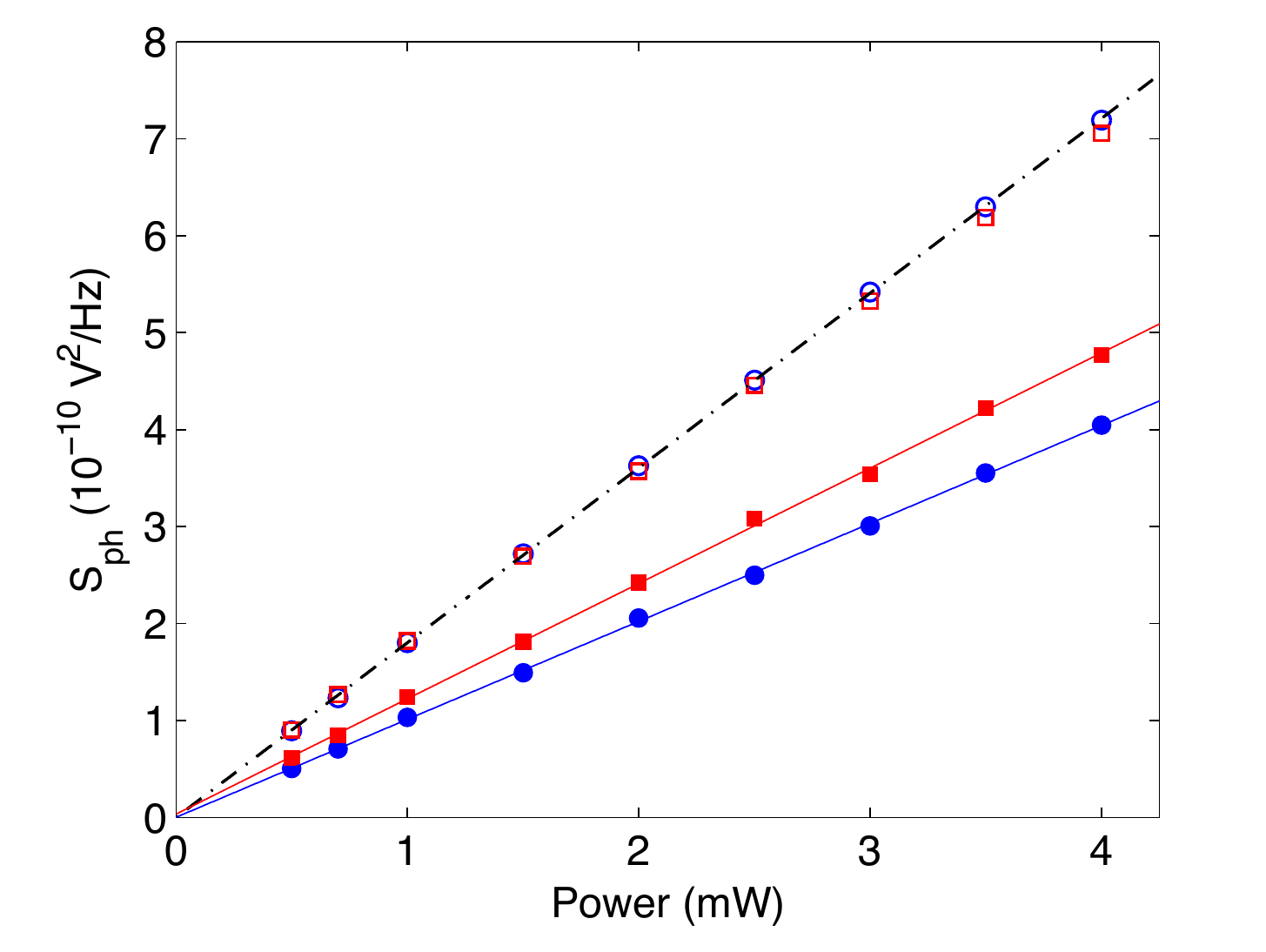}
    \caption{Shot-noise background, $S_{\rm ph}$, as a function of probe power reaching the detector for coherent probe (hollow symbols) and squeezed probe (filled symbols) at densities of
    $2.4 \times 10^{12} cm^{-3}$ (blue) and $9.3 \times 10^{12} cm^{-3}$ (red).}
    \label{fig:shot-noise}
\end{figure}

Figure \ref{fig:shot-noise} shows $S_{\rm ph}$, as estimated by fitting the measured PSD to Eq.~(1) in the main text, at different atomic densities for a coherent probe (hollow symbols) and squeezed probe (filled symbols). The dashed lines and solid lines in Fig.~\ref{fig:shot-noise} correspond to a fit of the data using Eq.~(\ref{Eq:Sph}), with $Q$ and $\xi^2$ as the free parameter in the fit for coherent-probe and squeezed-probe data, respectively. From the coherent-probe data, for which $\xi^2 = 1$, we obtain $Q =0.87$. The different slopes observed for the squeezing-probe data can be explained by the degradation of squeezing due to light absorption, given by \cite{Sorensen1998}

\begin{equation}
    \xi^2 = 1-(1-\xi^{2}_{0})\exp[-\mathrm{OD}],
    \label{Eq:squeezingfactor}
\end{equation}
where $\mathrm{OD}$ is the optical depth experienced by the light beam. For our experimental conditions $\xi^{2}_{0}= 0.55$, obtained by fitting the measured $\xi^{2}$ to Eq. (\ref{Eq:squeezingfactor}) and can be considered as the squeezing parameter of the transmitted light when the cell is at room temperature.

\subsection{Atomic noise } \label{section:AtomicNoise}

We can compute $\Theta_{\rm FR}$ by a coarse-grained approach. Dropping the $t$ for simplicity, and labelling by $i$ the isotope mass number and by $j$ the hyperfine state, so that $f^{(i,j)}$ is the single-atom total spin quantum number, the contribution to $\Theta_{\rm FR}$ from atoms in a small region of dimensions $\Delta x \times \Delta y \times \Lcell$, centered on $(x_{m},y_{m})$

\begin{equation}
 \Theta_{\rm FR}^{(i,j)} = \frac{1}{P}\frac{D_i(\nu')}{(2\INucl_i + 1)}
  \sum_{m} P(x_{m},y_{m}) \frac{\sigma_{0}}{\Delta x \Delta y} {F}_{z}^{(i,j,m)},
    \label{Eq:ThetaFFromFz}
\end{equation}
where $P(x_{m},y_{m})\approx \Delta x\Delta y \Intens(x_{m},y_{m})$ is the power of the beam in the given region, $2\INucl_{i}+1$ is a geometrical factor accounting for the hyperfine coupling between electronic spin ($S=1/2$) and nuclear spin ($I_{i}$) of the atom, so that $f^{(i,j)}=I_{i}+j$, $j\in\{-\frac{1}{2},+\frac{1}{2}\}$, and ${F}_z^{(i,j,m)}$ is the $z$-component of the collective angular momentum operator, i.e., the sum of the individual angular momenta ${\bf f}^{(i,j)}$ for atoms in the given region.

The on-resonance cross-section for the collision-broadened optical line is \cite{happer2010Book}

\begin{equation}
 \sigma_{0}=\frac{cr_{\rm e}f_\mathrm{{osc}}}{\Delta\nulight/2} =  2.4 \times 10^{-12} ~ \mathrm{cm^2},
\end{equation}
where $r_{\rm e} = 2.82 \times 10^{-13} $ cm is the classical electron radius, $f_{\mathrm{osc}}=0.34$ is the oscillator strength of the D$_1$ transition in Rb, and $c$ is the speed of light. The spectral factor is

\begin{equation}
D_i(\nu') = \frac{(\nu'-\nu_{j}')\Delta\nulight/2}{(\nu'-\nu_{j}')^2+(\Delta\nulight/2)^2},
\end{equation}
where $\nu'$ and $\nu'_{j}$, denote the probe optical frequency and optical resonance frequency, respectively,  and $\Delta\nulight$ represents the pressure-broadened FWHM of the optical transition. For the vapor cell used in our experiments $\Delta\nulight \approx $ \SI{2.4}{GHz} due to \SI{100}{Torr} of N$_{\mathrm{2}}$ buffer gas.

\begin{table}[t]
\centering 
\begin{tabular}{r r r} 
\hline\hline 
Parameter & Value  & Unit \\ [0.1ex] 
\hline 
$Q$ & $0.87$   & ---\\ 
$\alpha/(2\pi)$ & $57.8$ & $\rm{Hz}/(10^{12} cm^{-3})$ \\
$\beta/(2\pi)$ & $63$ &$\rm{Hz}/mW$  \\ 
$\Gamma_{0}/2\pi$ & $501$ & Hz \\
$\sigma_{0}$ & 2.4 $\times 10^{-12}$ & cm$^2$ \\
$\kappa^2$ & $5.0 \times10^{-4}$ & ---\\
$A_{\rm eff}$ &  0.0544 & cm$^2$ \\
$\Delta\nulight$ &  2.4 & GHz \\
\hline 
\end{tabular}
\caption{Parameters used in computing $\eta$.  See text for details.}
\label{table:parameters} 
\end{table}

For a given region, the mean of the collective spin projection is $\langle F_{z}^{(i,j,m)} \rangle = n_{i} \Delta x  \Delta y \Lcell {\rm Tr}[\rho {f}_{z}^{i,j}] = 0$ where $n_{i}$ is the atomic density of the $i$-th species, and $\rho$ is the thermal state, which to a very good approximation is a uniform mixture of the ground states. As a result $\langle\Theta_{\rm FR}^{(i,j)}\rangle=0$.

In a similar way, and assuming that different atoms are independent, so that their variances sum, we find
\begin{eqnarray}
 \var \Theta_{\rm FR}^{(i,j)}
 & = & \left(\frac{\sigma_{0}}{P}\frac{D_i(\nu')}{(2\INucl_i + 1)}\right)^2
  \var \sum_{m} \Intens(x_{m},y_{m})  {F}_{z}^{(i,j,m)}
 \nonumber \\ & \propto &
  \sum_{m}   \var  \Intens(x_{m},y_{m}) {F}_{z}^{(i,j,m)}
  \nonumber \\ & \propto &
   \sum_{m} \Intens^2(x_{m},y_{m})   \var  {F}_{z}^{(i,j,m)},
   \end{eqnarray}
where $\var F_{z}^{(i,j,m)} = n_{i}\,\Delta x\,\Delta y\,\Lcell {\rm Tr}[\rho ({f}_{z}^{(i,j)})^2]$, with
\begin{equation}
{\rm Tr}[\rho ({f}_{z}^{(i,j)})^2]  =  \frac{f^{(i,j)}(f^{(i,j)}+1)(2f^{(i,j)}+1)}{6(2\INucl_i+1)}.
\end{equation}

Taking the limit $\Delta x \Delta y \rightarrow dx\,dy$, and assuming the contributions of different isotopes and different hyperfine levels contribute independently, the spin noise due to isotope $i$ is
\begin{equation}
 \var \Theta_{\rm FR}^{(i)} = \kappa_{i}^2\sigma_{0}^2 n\Lcell \frac{\int dxdy \Intens^2(x,y)}{P^2},
    \label{Eq:VarTheta}
\end{equation}
where the parameter $\kappa_{i}^2$ is given by
\begin{eqnarray}
\kappa_{i}^2 &=& \sum_j
 \frac{D^2_i(\nu')}{(2\INucl_i + 1)^3}
\frac{ f^{(i,j)}(f^{(i,j)}+1)(2f^{(i,j)}+1)}{6}
           \label{Eq:kappa}
\end{eqnarray}

Equation (\ref{Eq:VarTheta}) is conveniently expressed as
\begin{equation}
 \var \Theta_{\rm FR}^{(i)} = N_{i} \frac{\sigma_{0}^2}{\Aeff^2}\kappa_{i}^2,
     \label{Eq:Intspn}
\end{equation}
where $N_{i} \equiv n_{i} A_{\rm eff}L_{\rm cell}$  is the effective number of istope-$i$ atoms in the beam, and $\Aeff$ is the effective area \cite{Shah2010}:
 \begin{equation}
 \Aeff \equiv \frac{\left[\int dx dy \, \Intens(x,y)\right]^2} {\int dx dy \, \Intens^2(x,y) }.
 \end{equation}

The spin noise oscillates at the Larmor frequency $\nu_i$ and with FWHM linewidth $\Delta\nu_i$, so that
\begin{equation}
    S(\nu) = S_{\rm ph} +  \sum_{i\in\{85,87\} } S_{\rm at}^{(i)} \frac{(\Delta \nu/2)^2}{ (\nu-\nu_i)^2 + (\Delta\nu/2)^2}
    \label{Eq:Sat}
\end{equation}
where
\begin{equation}
    S_{\rm at}^{(i)}(\nu) = \frac{4G^2\Re^2P^2}{\pi \Delta\nu/2} \var \Theta_{\rm FR}^{(i)}.
    \label{Eq:Sat}
\end{equation}

\begin{figure}[t]
    \centering
    \includegraphics[scale=0.45]{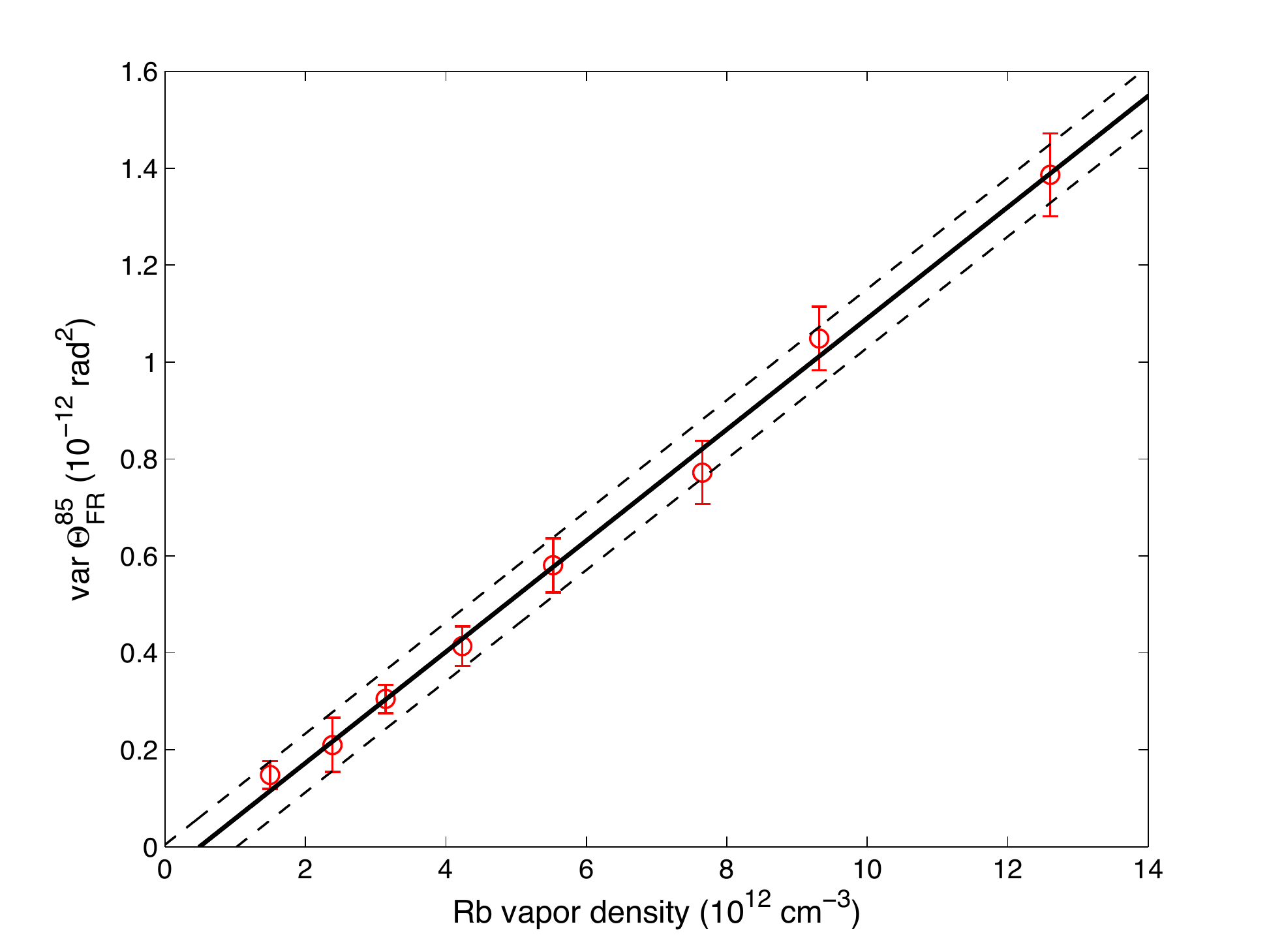}
    \caption{$\var \Theta_{FR}^{(85)}$ as a function of atomic density for coherent probe. The solid line shows to a linear fit to the data.  Dashed lines show the best fit $\pm 3\sigma$ statistical uncertainty in the fit offset.}
    \label{fig:spin-noise}
\end{figure}

From the fitted amplitude $S_{\rm at}^{(85)}$, and FWHM $\Delta \nu_{85}$ of the Rb$^{85}$ spin noise spectrum  we compute $\var\Theta_{\rm FR}^{(85)}$ using Eq.~(\ref{Eq:Sat}). These data are shown in Fig.~\ref{fig:spin-noise} as a function of the Rb-vapor density $n$. The solid line in Fig.~\ref{fig:spin-noise} corresponds to a fit of the data using Eq.~(\ref{Eq:Intspn}) with $\kappa^2_{85}$ as the free parameter and with $N_{85}=0.72~nA_{\mathrm{eff}}L_{\mathrm{cell}}$, here  $A_{\rm eff}= \rm 0.054 ~cm^{2}$ and $L_{\rm cell}= 3 \rm ~ cm$. From the fit we obtain $\kappa_{85}^{2} =5 \times 10^{-4}$, to be compared with the value of \num{3.7d-4} obtained by evaluating Eq.~(\ref{Eq:kappa}) with a detuning of $\nu'-\nu'_{85} = \SI{- 20}{GHz}$  and optical linewidth $\Delta \nulight$ = \SI{2.4}{GHz}, and using $\sigma_{0} = 2.4 10^{-12} cm^2$.

\subsubsection{Width of the magnetic line} \label{subsection: spin relaxation data}

For $^{85}$Rb the FWHM width of the magnetic line is  $\Delta \nu = 1/(\pi T_{2})$, and can be approximated by  \cite{Jimenez2012}

 \begin{equation}
    \frac{1}{T_{2}} = \Gamma_{0} + \alpha n + \beta P,
    \label{Eq:T2SI}
\end{equation}
where $\alpha n$ and $\beta P$ are the contributions from atomic collisions and power-broadening, respectively, with $n$ being the Rb-vapor density, $P$ the optical power of the probe beam, and $\Gamma_{0}$ the spin relaxation due to other mechanisms, including buffer-gas collisions and the finite residence time of the atoms in the light beam.

To estimate the power broadening and collisional-broadening parameters we fit the measured $\Delta \nu_{85}$ using Eq.~(\ref{Eq:T2SI}). From the fit we obtain a collisional broadening parameter $\alpha/(2\pi) = 57.8$Hz/$10^{12}cm^{3}$, a power-broadening parameter $\beta/(2\pi) = 63$Hz/mW, and $\Gamma_{0}/(2\pi) = 501$ Hz.

\subsection{SNR}

Using equations (\ref{Eq:Sph}), (\ref{Eq:Intspn}), (\ref{Eq:Sat}), and (\ref{Eq:T2SI}) we obtain the following expression for the SNR $\eta_{i} \equiv \Sat^{(i)}/\Sph$
\begin{equation}
    \eta_i = \frac{P}{\hbar \omegalight} \frac{4 Q }{\xi^2} \frac{\sigma_{0}^2}{A_{\mathrm{eff}}^2}\frac{\kappa_{i}^2N_{i}}{\Gamma},
\end{equation}
where $N_{i}=n_{i}A_{\mathrm{eff}}L_{\mathrm{cell}}$, thus arriving to Eq.~(2) of the main text.

\bibliographystyle{apsrev4-1}
\bibliography{SqueezedSNS}

\end{document}